\documentclass[12pt]{article}
\usepackage[margin=1in]{geometry}
\usepackage{graphicx}
\usepackage{subcaption}
\graphicspath{{./plots/}} 
\usepackage{float}
\usepackage{amsmath}
\usepackage[none]{hyphenat}
\usepackage[numbers]{natbib}
\usepackage{amssymb}
\usepackage{url} 

\usepackage{amsthm}

\usepackage[title,titletoc]{appendix}

\usepackage{enumitem}

\usepackage[none]{hyphenat}

\makeatletter
\g@addto@macro\bfseries{\boldmath}
\makeatother

\usepackage[bottom]{footmisc}

\usepackage{anyfontsize}

\newcounter{desccount}

\newcommand{\descref}[1]{\hyperref[#1]{#1}}

\usepackage{hyperref}

\begin{document}
	
	\title{A Study of UK Household Wealth through Empirical Analysis and a Non-linear Kesten Process}
	\author{Samuel Forbes\thanks{S.Forbes.1@warwick.ac.uk} \and Stefan Grosskinsky\thanks{S.W.Grosskinsky@tudelft.nl}}
	\date{July 2021}

	\maketitle
	
	\begin{abstract}
		We study the wealth distribution of UK households through a detailed analysis of data from wealth surveys and rich lists, and propose a non-linear Kesten process to model the dynamics of household wealth. The main features of our model are that we focus on wealth growth and disregard exchange, and that the rate of return on wealth is increasing with wealth. The linear case with wealth-independent return rate has been well studied, leading to a log-normal wealth distribution in the long time limit which is essentially independent of initial conditions. We find through theoretical analysis and simulations that the non-linearity in our model leads to more realistic power-law tails, and can explain an apparent two-tailed structure in the empirical wealth distribution of the UK and other countries. Other realistic features of our model include an increase in inequality over time, and a stronger dependence on initial conditions compared to linear models.
	\end{abstract}
	
	\subsection*{Acknowledgements}
	We would like to thank Alexander Karalis Isaac and Colm Connaughton for their helpful discussions on this work. S. Forbes would also like to acknowledge financial support from EPSRC through grant EP/L015374/1.


	\newpage
	
	\section{Introduction}
	
	The dynamics of wealth and income inequality is a subject of increasing research interest and public debate, encapsulated by major works such as Piketty's `Capital in the 21st Century' \cite{piketty2013capital}. The recent COVID-19 pandemic has added to the debate on inequality as some of the very richest, particularly in the tech industry, have gained large quantities of wealth whilst many in the population have faced redundancies and reliance on government benefits \cite{berkhout2021inequality}. Data on standard inequality measures, such as the Gini coefficient or wealth/income shares, clearly indicate that inequality has increased since the 1980s in many areas of the world \cite{alvaredo2018top}. Potential contributing factors include globalisation, financialisation, decreased taxes, increased tax evasion and avoidance, increased inheritance and domination of the technological sector \cite{piketty2013capital, alvaredo2018top, atkinson2018wealth}. In this paper we summarise these multitude of factors into an idealised growth model for household wealth, dominated by one simple effect: that the wealthier you are, the higher  your rate of return (ROR), i.e. the return on wealth you are likely to receive grows superlinearly with wealth. We refer to this type of reinforcement dynamics in our discrete time model as a non-linear Kesten process, which is a generalisation of the work on linear reinforcement initiated by Kesten  \cite{kesten1973random}. The increasing dependence of RORs on wealth has been confirmed in recent studies \cite{fagereng2020heterogeneity, bach2016rich, ederer2020rich}, and we present further empirical evidence for the UK.
	
	Stochastic models with multiplicative noise applied to income and wealth dynamics have a long history in economics, with an early
	major publication in 1953 by Champernowne \cite{champernowne1953model}, and since then have been applied extensively and are summarised in several reviews, see for example \cite{gabaix2009power, benhabib2018skewed}. These models have been used as they exhibit power-law tails, which is a key feature of both income and wealth distributions. Research in the field of Econophysics has focused mostly on exchange of money or wealth, in analogy to energy transfer in models of statistical mechanics (see e.g.\ \cite{yakovenko2009colloquium, chatterjee2007econophysics} for an overview). It has been found also in this context, that additive noise leads to Boltzmann-Gibbs type distributions with exponential tails, and heavy tails can result from multiplicative noise or disorder
	\cite{chatterjee2004pareto}.
	The focus on pure exchange dynamics has been recognised as unrealistic to model wealth (see \cite{yakovenko2009colloquium} page 13),
	but only very few studies consider both exchange and growth.
	In \cite{berman2020wealth, bouchaud2000wealth} the authors study growth dynamics of wealth with a global redistribution dynamics, inducing a weak mean-field type interaction between agents. 
	In our model we disregard wealth exchange between households and focus entirely on growth dynamics. This is of course a simplification, but in our view and in line with previous studies mentioned above, growth is clearly the dominant aspect of wealth dynamics for most households, and on average 
	nominal wealth has been growing in an exponential fashion since at least the industrial revolution \cite{mclaughlin2014historical}.

\begin{figure}
		\centering
		\captionsetup{justification=centering}
		\includegraphics[scale=0.35]{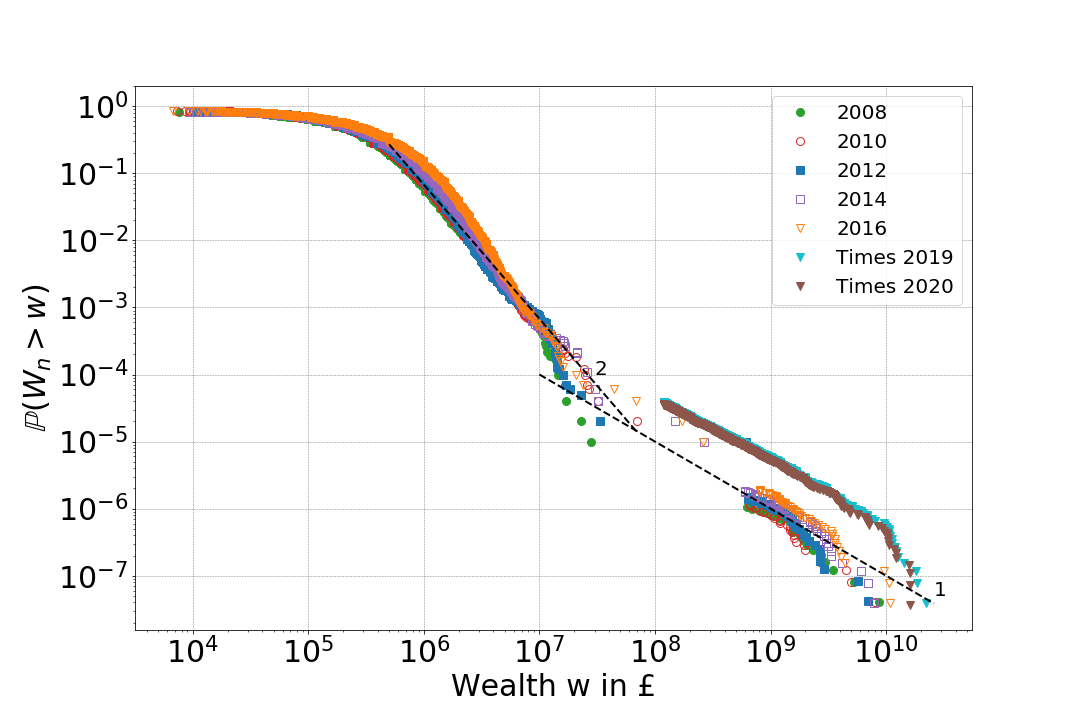}
		\caption{Empirical tail distribution of positive UK household wealth for five consecutive time periods 2008, 2010, 2012, 2014, 2016 from the ONS wealth and asset survey \cite{was}, together with Forbes rich list data on billionaires \cite{Forbes_rl}, and UK Times rich list data from 2019, 2020 (see Appendix \ref{sec_data}). Dashed lines indicate power-law tails with exponents $2$ and $1$ for comparison.}
		\label{survey_times_forbes2}
	\end{figure}

	Figure \ref{survey_times_forbes2} shows the tail of the household wealth distribution for the UK from recent wealth and asset survey data \cite{was} and rich lists \cite{Forbes_rl,Times2020_rl}. We see here the presence of two power laws in the upper tail: one for the richest in the survey with exponent around $2$, and one for the richest in society found in the rich lists with exponent around $1$. Such a change in power-law exponent has been observed for other countries and argued to be a generic sampling artefact from survey bias in the data \cite{vermeulen2018fat}. However due to the particular strength of the effect we believe that the two-tailed structure is a genuine feature of the data. 
	From previous studies \cite{hitczenko2011renorming} linear Kesten processes are known to lead to asymptotic log-normal distributions of wealth. 
	Our non-linear model produces a power-law tail from various generic initial conditions, and in the long run also a two-tailed structure due to a crossover phenomenon resulting from the non-linearity, which we will explain in detail.

	We also find that our model has a strong dependence on initial conditions, corresponding to the idea of a low social mobility \cite{berman2018long}. It is particularly suitable to describe wealth dynamics since the 1980s, when deregulation of financial markets started to facilitate increasing rates of return for assets typically held by wealthier agents \cite{balder2018financialization}, providing increased access to credit and investment opportunities. 
	During the 2007-2008 financial crisis, shortage of available credit temporarily also affected wealth growth for households \cite{bricker2020wealth}. But after a relatively short period of adaption and in spite of declining interest rates \cite{schmelzing2019eight}, prices of e.g.\ housing and financial assets are again increasing at close to pre-crisis levels \cite{blakeley2021financialization}, so the main premise of our model remains valid. 
	While an important macroeconomic question, the mechanisms behind wealth growth are not part of our discussion and we focus on the distribution of wealth among households.
	Throughout this paper we only model positive wealth, while appreciating that a significant fraction (above 10\% \cite{was}) of the UK population has negative wealth, i.e.\ is in debt. This requires additional modelling and the dynamics we propose do not apply in this case.
		
	Wealth can be defined as assets minus liabilities \cite{Atkinson_U_S} and is usually measured in a particular currency, GBP in our case. It can be interpreted as the balance sheet of a household, and therefore only assets that can be assigned a monetary value contribute, excluding e.g.\ health or education of members of a household. 
We also note that wealth is a stock of value unlike income, which represents a flow of value over time. 
	The wealth and asset surveys categorise wealth into four components: physical, financial, property and pension \cite{ONS_wealth_2008}.
	In our model we assume that wealth increases on average due to two mechanisms: multiplicative growth due to returns on current wealth, and additive savings such as excess salary that is not spent on living costs and other expenses which do not contribute to the balance sheet of the household. 
	In general, wealthier agents can diversify their assets, including riskier strategies with higher average returns \cite{bach2016rich, ederer2020rich}. Different composition of wealth in different wealth deciles is provided in the asset survey data \cite{was} and summarised in Appendix \ref{sec_returns}.

		

	\section{Model}
	\label{sec_model}
	
	We consider independent agents (representing households), whose wealth at discrete time $n \in \{0,1,2,\dots\}$ (representing years) is denoted by $W_n \geq 0$. As explained in the introduction, we focus on wealth growth rather than exchange, and model the dynamics of positive wealth only, keeping track of bankruptcy events after which we reset the wealth value of the agent (see Section \ref{sec_sim} for details). We assume that the wealth of an agent over the time period $n$ to $n+1$ changes via two mechanisms: \textbf{returns on existing wealth}, where $R_{n+1} \in\mathbb{R}$ denotes the corresponding rate of return (ROR), and \textbf{savings} $S_{n+1} \geq 0$, resulting for example from excess earnings which are independent of the current wealth of an agent (see Section \ref{savings} for details). This leads to the recursion
	\begin{equation}
	W_{n+1}=W_n(1+R_{n+1})+S_{n+1}\quad\mbox{with initial condition}\quad W_0 >0.
	\label{initw_eq}
	\end{equation}
	Here the RORs $R_n$ and savings $S_n$ are independent random variables. It is commonly accepted that RORs depend monotonically on wealth \cite{fagereng2020heterogeneity, bach2016rich, ederer2020rich}, and we assume the following power-law form,
	\begin{equation}
	R_{n+1} = \alpha_{n+1}W_n^{\gamma-1}\quad\mbox{for some }\gamma \geq 1,
	\label{R_n}
	\end{equation}
	where $\alpha_n \in\mathbb{R}$ are i.i.d.\ random variables from some fixed probability distribution, and with small probability can also take negative values. 
	The very simple choice \eqref{R_n} is consistent with empirical data for the UK presented in Section \ref{sec_returns_main}. We are not claiming that this is the best or most detailed model for RORs, which have been observed in some cases to exhibit an intermediate plateau rather than a strict increase as a function of $W_n$ (see e.g. Fig. 2 in \cite{ ederer2020rich}). But our aim here is to capture the most essential features in a simple model that can also be analysed analytically, and it is of course possible for simulations to replace \eqref{R_n} by different functions. 
	We find that a non-central $t$ distribution (see Appendix \ref{sec_nct} for details)  provides a good match with data for $\alpha_n$, which is discussed in Section \ref{sec_empirics}, Figure \ref{alpha_bil_was_kd}.
	
	Substituting (\ref{R_n}) in (\ref{initw_eq}) gives the recursion
	\begin{equation}
	W_{n+1} = W_n+\alpha_{n+1}W_n^{\gamma}+S_{n+1}.
	\label{nlkp}
	\end{equation}
	With $\gamma>1$ we refer to (\ref{nlkp}) as a \textbf{non-linear Kesten process}. We now summarise theoretical results of \eqref{nlkp} for different $\gamma$ values.\\
	
	\textbf{$\gamma =1$.} In this case $R_n =\alpha_n$ and $W_{n+1} =(1+\alpha_{n+1} ) W_n +S_{n+1}$. The stationary version of this linear model has been introduced and studied by Kesten \cite{kesten1973random}, and the non-stationary asymptotic growth case is more recently discussed in \cite{hitczenko2011renorming}. It is easy to see that the asymptotic behaviour of $W_n$ is dominated by the exponential $e^{n\log |1+\alpha_n |}$, and we present details on the analysis of both cases in Appendix \ref{ap_lkp}. In the stationary case with $\mu :=\mathbb{E} [\log |1+R_n |] <0$, the model is known to exhibit power-law tails in the limiting distribution, but for wealth dynamics the non-stationary case of asymptotic growth is most relevant, which occurs for $\mu >0$. 
	Following results in \cite{hitczenko2011renorming}, the asymptotics is given by a \textbf{log-normal distribution} such that to leading exponential order\footnote{Here the symbol $\asymp$ means that $W_n = W_0 \exp\Big( \mu n+\sqrt{n\nu^2 } Z +o(\sqrt{n})\Big)$ as $n\to\infty$, with Bachmann-Landau (or little $o$) notation such that $o(a_n )/a_n \to 0$ for all positive sequences $(a_n :n\in\mathbb{N} )$. }
	\begin{equation}\label{linres}
	W_n \asymp W_0 \exp\Big( \mu n+\sqrt{n\nu^2 } Z\Big)\quad\mbox{as }n\to\infty,
	\end{equation}
	where $\nu^2 :=\mathrm{Var}[\log |1+R_n |]$ and $Z \sim \mathcal{N}(0,1)$ is a standard Gaussian. The rigorous version of this result is subject to further reasonable and mild regularity assumptions on the distributions of parameters (see Theorem 2(i) in \cite{hitczenko2011renorming}), and the leading order behaviour is independent of the savings $S_n$. Since \eqref{nlkp} is linear in $W_n$, the model also has a natural scale invariance for the units of wealth (see discsussion in \cite{bouchaud2000wealth}), and the initial condition $W_0$ enters \eqref{linres} as a simple multiplicative constant.\\ 
	
	\textbf{$\gamma >1$.} To our knowledge the non-linear model has not been studied before. Details are given in Appendix \ref{ap_nlkp}, where we find asymptotic super-exponential growth to leading order,
	\begin{equation}
	W_n \asymp \big( W_0 e^D \big)^{\gamma^n}\quad\mbox{as }n\to\infty,
	\end{equation}
	where $D$ is given by a convergent series depending on the distribution of $\alpha_n$ and the initial behaviour of the process. Again, we focus on the non-stationary case with $W_0 e^D >1$. In contrast to the linear case, we see that the asymptotics depend in a strong, non-linear way on the initial conditions and early dynamics of the process. Therefore there is no central limit theorem on the logarithmic scale that leads to \eqref{linres}, and we are not able to predict the asymptotic scaling distribution of $W_n$. But numerical results presented in Section \ref{sec_sim} show that the model exhibits power-law tails with realistic shapes on relevant time scales.
	
	For realistic initial conditions and parameters the dynamics follows initially an exponential growth regime, and super-exponential growth sets in when the dominant term in yearly gains in equation (\ref{nlkp}) changes from $W_n$ to $\alpha_{n+1} W_n^\gamma$ (additive savings again do not influence the asymptotic behaviour). This means that the returns from wealth in a single year become of the same order or higher than current wealth, which happens for values around
	\begin{equation}\label{crossover}
	W_n \approx \alpha_{n+1}^{-1/(\gamma-1)}.
	\end{equation}
Billionaire return data in Figure \ref{returns_all2} below indeed confirm that RORs of around $100\%$ or more can be achieved. From numerical results in Section \ref{sec_sim} we see that this crossover leads to a two-tailed structure of the distribution of $W_n$ similar to what we see in the data in Figure \ref{survey_times_forbes2}, and we think this feature of the model provides a promising explanation for this effect. Since we find in the next section that $\gamma$ is close to $1$, \eqref{crossover} is very sensitive to the value of the random variable $\alpha_{n+1}$ (which is raised to a large power), leading to a broad crossover region.  While this crossover is a realistic feature seen in data from the UK and other countries (\cite{vermeulen2018fat}, but notably not in the USA, see online Appendix of \cite{vermeulen2018fat}), the non-linearity also implies that the model is not scale invariant and coefficients will heavily depend on the currency unit.
	
	We further find empirically that $\alpha_n$ is mostly positive with a heavy tail, but negative values are possible, see Figure \ref{alpha_bil_was_kd} of Section \ref{sec_pref_alpha}, and thus $W_n$ may become negative. Since our dynamics \eqref{initw_eq} are not built to describe agents in debt, we replace 
	$W_n$ with one of three replacement mechanisms discussed in Section \ref{sec_gen_sim}.
	We note that bankruptcy events where agents' losses exceed their current wealth are realistic and do occur, but in this paper we focus on modelling the dynamics of agents with positive wealth.\\

	We also note that both, the non-stationary linear and super-linear models, exhibit \\ \textbf{monopoly}, where the wealth fraction of the richest agent in a system of $N$ independent agents tends to $1$ as time $n\to\infty$. This behaviour is well known for distributions with heavy tails (see e.g. Table 3.7 in \cite{goldie1998subexponential}), which include the log-normal distribution in the linear case \eqref{linres}, and is only more pronounced in the super-linear model with heavier tails. We present related numerical results for the Gini coefficient and the top 1\% wealth share in simulations, both tending to $1$ in the long-time limit. While of course this extreme limit is not realistic currently, inequality measures are well known to increase since the 1980s (see summary in Appendix \ref{sec_ineq}). This is consistent with understanding current wealth distributions as transient behaviour of our model, which leads to monopoly if parameters remain unchanged over time. Of course we can only parametrise our model over the current range of wealth values, and in order to get more realistic forecasts for future wealth distributions, we would have to include also the lifetime and inheritance dynamics for agents and the role of external influences (such as war or other catastrophies). The simplified model we present here explains how current wealth distributions can arise naturally from generic initial conditions, and we discuss possible refinements for further study in Section \ref{sec_discussion}.
	
	\section{Empirics}
	
	\label{sec_empirics}
	
	Before moving on to the simulations of the non-linear Kesten process (\ref{nlkp}) we undertake some key empirical analysis to parametrise the model. We calculate returns on wealth, $R_n$, and the prefactor, $\alpha_n$, and make statistical fits on these variables. Although savings do not evolve with wealth as mentioned above, they are correlated with initial wealth values of an agent as part of their social status or fitness. To infer this dependence, we look at UK income and expenditure data for the year 2016 \cite{ONS_income, ONS_expenditure}. 
	
	\subsection{Returns $R_n$} 
	
	\label{sec_returns_main}
	
	From (\ref{nlkp}) we rearrange to find the ROR as 
	
	\begin{equation}
	R_{n+1} = \frac{W_{n+1}-W_{n}-S_{n+1}}{W_n} \approx \frac{W_{n+1}-W_{n}}{W_n}\quad\mbox{for billionaires.}
	\label{returns_eq}
	\end{equation}
	For wealthy agents, wealth gain is to a large extent dominated by returns on wealth, so that $W_{n+1} -W_n \gg S_{n+1}$ and savings can typically be ignored. The ROR is then simply given by the wealth growth rate, which we will use to compute $R_n$ for billionaires, while we include savings to estimate ROR from survey data for other agents.
	
\begin{figure}[H]
		\centering
		\captionsetup{justification=centering}
		\includegraphics[scale=0.35]{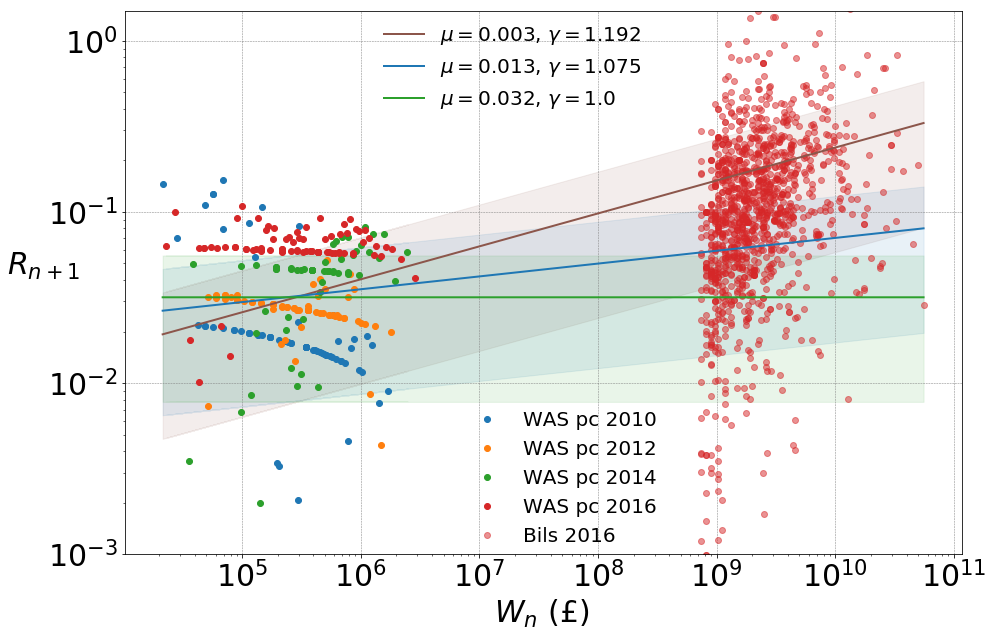}
		\caption{Percentile ROR using WAS data \cite{was} for the years 2010, 2012, 2014 and 2016, and ROR for individual billionaires for 2016 \cite{Forbes_rl}. Power law fits according to \eqref{mudef} to the cluster of WAS ROR data combined over all four time periods, leads to $\mu \approx 0.003$, $\gamma \approx 1.192$ (with both parameters free) and to $\mu \approx 0.013$ with chosen $\gamma = 1.075$ (justified below in Figure \ref{alpha_bil_was_kd}). We also include $\gamma =1$ for comparison, leading to $\mu \approx 0.032$, i.e.\ an average ROR of about $3\%$. Respective shaded regions are one standard deviation around the power fit means (\ref{one_std_mean}) as explained in Appendix \ref{ap_mean_var_ret}. }
		\label{returns_all2}
	\end{figure}

\begin{figure}[H]
		\centering
		\captionsetup{justification=centering}
		\includegraphics[width=\textwidth]{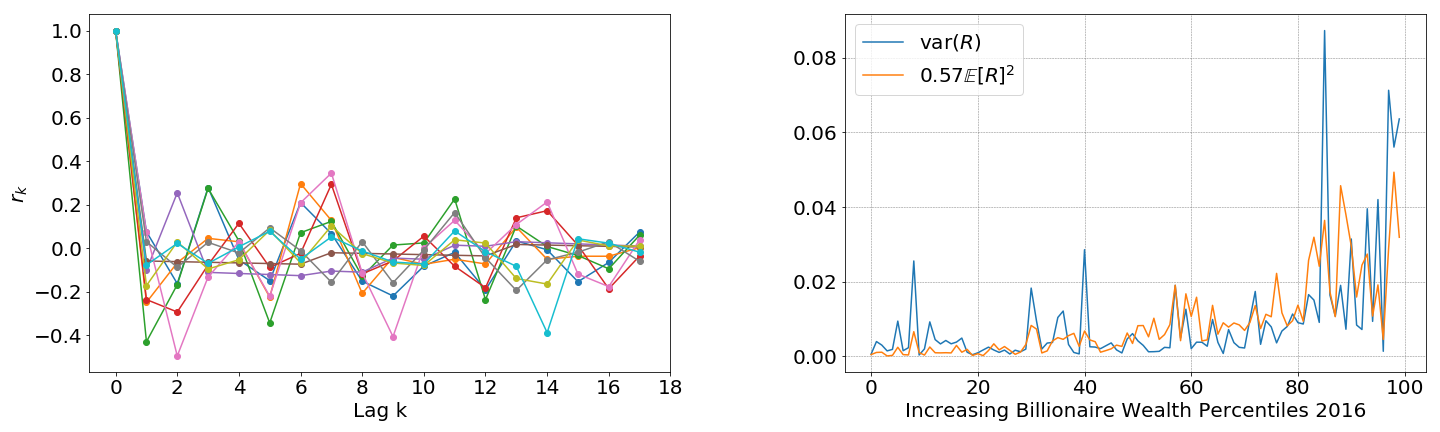}
		\caption{Left: autocorrelation of a sample of billionaire ROR indicating independence in returns. Right: average annual billionaire returns from 2008-2016 Forbes list \cite{Forbes_rl}, showing mean and variance relationship for increasing wealth percentiles as in \eqref{meanvar}.}
		\label{returns_bil_mean_var_aut}
	\end{figure}

	As mentioned previously, fairly recent work \cite{fagereng2020heterogeneity, bach2016rich, ederer2020rich} has suggested an increasing wealth dependence on returns. We also find empirical evidence for this from UK survey data as summarised in Figure \ref{returns_all2}, and assume a simple power-law relationship as in  (\ref{R_n}) which is roughly consistent with the data. According to this we have
	\begin{equation}\label{mudef}
	    \mathbb{E}[R_{n+1}|W_n] = \mu W_n^{\gamma-1}\ ,\quad\mbox{where}\quad \mu =\mathbb{E}[\alpha_{n+1}].
	\end{equation}
	We fit the power-law exponent $\gamma$ and the prefactor $\mu$ as shown in Figure \ref{returns_all2}, and also find evidence that returns are independent across time and the variance of returns is proportional to the square of the mean returns as wealth increases (see Figure \ref{returns_bil_mean_var_aut}),
	\begin{equation}\label{meanvar}
	    \text{var}(R_{n+1} |W_{n}) \approx 0.57 \,\mathbb{E}[R_{n+1} |W_{n}]^2.
	\end{equation}
	Such a quadratic scaling relationship of mean and variance is common in multiplicative processes, and consistent with our model assumption \eqref{R_n}, as is explained in Appendix \ref{ap_mean_var_ret}.
	
	Note that the apparent structure in percentile return data in Figure \ref{returns_all2} for individual years does not constitute reliable information in our view, since the variation of the points is artificially decreased due to our numerical procedure as explained in \ref{sec_returns}. Viewing all years as a combined dataset, we find an increasing wealth dependence of RORs consistent with a simple power-law relationship, which also matches well with data for billionaires. In the next subsection we present a method to estimate a reasonable value of the power-law exponent $\gamma$ so that both, WAS and billionaire return data, can modelled well with our assumption on returns \eqref{R_n}.
	%

	\subsection{Fitting $\alpha_n$}
	\label{sec_pref_alpha}
	
\begin{figure}[H]
		\centering
		\captionsetup{justification=centering}
		\includegraphics[width=\textwidth]{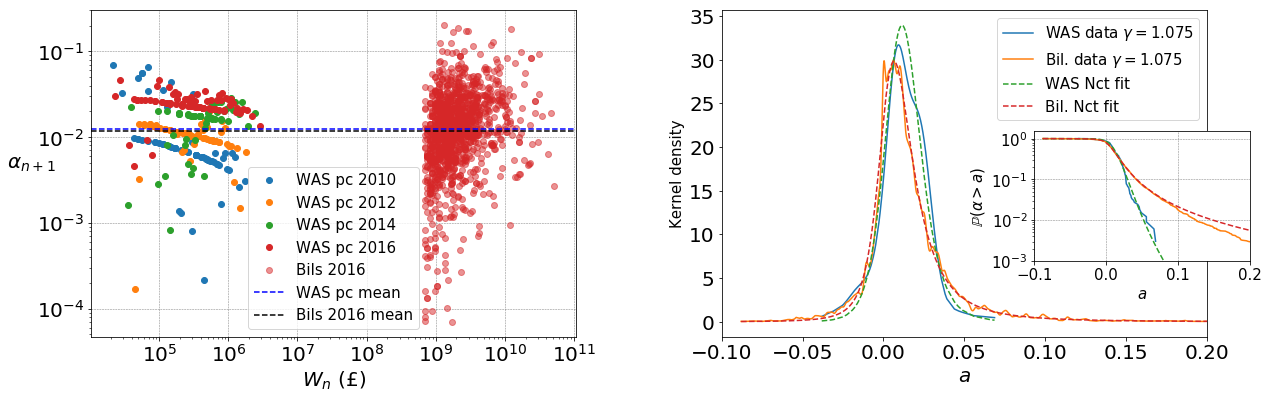}
		\caption{Left: $\alpha_{n+1}$ (\ref{alpha_n}) for WAS data  percentiles \cite{was} for four time periods along with 2016 billionaire data plotted against wealth $W_n$. We choose $\gamma =1.075$ so that the means of WAS and billionaire data essentially agree (dotted lines). 
		Right: Kernel density of $\alpha_{n+1}$ for WAS data and 2016 billionaire data as seen in the left Figure. Inset: corresponding empirical tails $\mathbb{P}(\alpha_n>a)$ on logarithmic scale. Dotted green and red lines provide fits by the non-central $t$-distribution (nct) to WAS and billionaires with respective nct parameter fits $k\approx6.03$, $c \approx 0.0573$, $l \approx -0.00575$, $s \approx 0.0112$ and $k\approx2.01$, $c \approx 0.941$, $l \approx -0.00156$, $s \approx 0.0112$.}
		\label{alpha_bil_was_kd}
	\end{figure}

	With (\ref{nlkp}) we have in analogy to \eqref{returns_eq} 
	
	\begin{equation}
	\alpha_{n+1} = \frac{W_{n+1}-W_{n}-S_{n+1}}{W_n^{\gamma}}\approx \frac{W_{n+1}-W_{n}}{W_n^{\gamma}} \quad\text{ for billionaires}.
	\label{alpha_n}
	\end{equation}
	As illustrated in Figure \ref{alpha_bil_was_kd}, we choose the power-law exponent $\gamma =1.075$, such that the return data from WAS and billionaires can be best explained with a single power law of the form \eqref{R_n}. 
	We fit the distribution of the $\alpha_n$ (which we assume to be i.i.d.) with a shifted and scaled non-central $t$-distribution (nct), i.e. we take
	\begin{equation*}
	\alpha_n \sim \text{nct}(k,c,l,s).
	\end{equation*}
	This distribution has four parameters: $k>0$ represents the degrees of freedom controlling the heaviness of the tail, $c\in \mathbb{R}$ is the centrality that controls the skewness of the distribution, $l\in\mathbb{R}$ is the shift and $s>0$ is the scale, see Appendix \ref{sec_nct} for details.

	We find that, while the bulk of the distributions of $\alpha_n$ agree well, the billionaire data lead to heavier tails than WAS data. Again, our method of extracting returns from WAS data leads to decreased fluctuations, and therefore we use the parameter values corresponding to billionaire data in simulations in Section \ref{sec_sim}.

	\subsection{Savings $S_n$}
	\label{savings}
	We recall that in our model \eqref{initw_eq} savings $S_n$ represent all contributions to wealth growth that are independent of the current wealth of an agent. They do not evolve with increasing wealth and only contribute additive noise, which does not influence the long-time behaviour of the dynamics. However, we need to estimate savings and their correlation with (initial) wealth to run simulations, and in particular in order to extract empirical RORs from wealth data using \eqref{returns_eq}, which determine the statistics of the crucial parameter $\alpha_n$. \cite{keister2017double} presents evidence for recent years in the US, that income and salary are positively correlated with wealth.
	
\begin{figure}[H]
		\centering
		\captionsetup{justification=centering}
		\includegraphics[width=\textwidth]{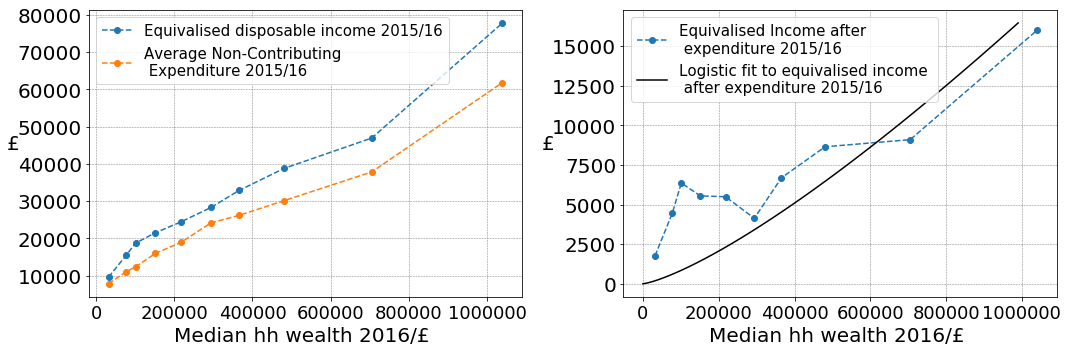}
		\caption{
		Left: Plot of equivalised disposable income and average household expenditure for 2015/16 against 2016 median wealth deciles. Right: Fit of the logistic function \eqref{snfit} to equivalised disposable income after expenditure, where we choose $\kappa_1 =10^6$ and fit $\kappa_2 =4.13\cdot 10^9$ and $\kappa_3 =-1.308$. ONS data sources used can be found in \cite{ONS_expenditure, ONS_income}.}
		\label{income_exp_wealth}
	\end{figure}

	We estimate savings by equivalised disposable income after expenditure for increasing deciles of median wealth using ONS data sources  \cite{ONS_expenditure, ONS_income}. 
	Equivalised disposable income is household size adjusted income available for spending after tax and deductions, and by expenditure we summarise costs that do not contribute to wealth, such as buying food or paying rent. 
	We fit the dependence on wealth $w$ with a logistic function
	\begin{equation}\label{snfit}
	    S(w)=\frac{\kappa_1}{1+\kappa_2 w^{\kappa_3}}\quad\mbox{with parameters }\kappa_1 ,\kappa_2 >0\mbox{ and }\kappa_3 <0.
	\end{equation}
	This is illustrated in Figure \ref{income_exp_wealth}, where we show data on equivalised disposable income, household expenditure and give the fitted parameter values for \eqref{snfit}. 
	
	
	We used \eqref{snfit} as an estimate for additive contributions to wealth growth when calculating percentile returns in Figure \ref{returns_all2}, see Appendix \ref{sec_returns} and in simulations in Section \ref{sec_real_init_condns} as a function of initial wealth $w=W_0$. Note that the logistic fit levels off at $\kappa_1 =10^6$ for large values of $w$ which is an arbitrary cap of $10^6$ GBP on wealth independent savings.
	For most rich households, contributions to wealth growth significantly beyond this scale are in the form of wealth returns.  It is important to note that none of our results are sensitive to the choice of parameters $\kappa_1$, $\kappa_2$ and $\kappa_3$, since savings only really play a role in parameter estimation or simulations on the scales shown in Figure \ref{income_exp_wealth}.

	\section{Simulation Results}
	\label{sec_sim}

For all simulations presented in this section we use i.i.d. $\alpha_n \sim \mathrm{nct}(k,c,l,s)$ with parameters
\begin{equation}\label{params}
k =2.008\ ,\quad c = 0.941\ ,\quad
l =-0.00156\quad\mbox{and}\quad s = 0.0112\ ,
\end{equation}
corresponding to data from individual billionaires which represent our best estimate of fluctuations for individual households for $\gamma=1.075$. We do, however, experiment with changing $\gamma$ values in which case we multiply the $\alpha_n$ by a positive constant to keep the mean at the same level. This is explained further in Section \ref{sec_gen_sim}.

	\subsection{Generic Initial Conditions with Zero Savings}
	
	\label{sec_gen_sim}
	
	To investigate the general properties and dependence on initial conditions of our model over longer time horizons, we consider the following four different initial conditions each with mean $10000$:
	
	\begin{enumerate}[leftmargin=1.9cm,label=\textbf{I.\arabic*}]
		\item $W_0=10000$  \label{init1} (BLUE)
		\item $W_0 \sim 5000+\text{Exp}(1/5000)$ \label{init2} (ORANGE)
		\item $W_0 \sim \text{Exp}(1/10000)$ \label{init3} (GREEN)
		\item $W_0 \sim \text{Pareto}(5000,2)$ \label{init4} (RED)
	\end{enumerate}
	In other words, in \ref{init1} all agents start with initial wealth $10000$, in \ref{init2} agents get $5000$ plus an exponentially distributed random amount with mean $5000$, in \ref{init3} initial wealth is drawn from an exponential with mean $10000$ and in \ref{init4} it is Pareto distributed with parameters $x_m=5000$ and exponent $\beta=2$.
	
	It is also possible in our simulations for the wealth $W_n (i)$ of an agent $i$ to become negative. 
	In this case we choose one of the following replacements for $W_n(i)$:
	
	\begin{enumerate}[leftmargin=2cm, label=\textbf{R.\arabic*}]
		\item replace with a proportion of the agent's previous positive wealth value \newline $p W_{n-1}(i)>0$ such that $p$ is uniformly chosen from $(0,1]$ \label{r1}
		\item replace with the agent's previous positive wealth value $W_{n-1}(i)>0$
		\label{r2}
		\item replace with wealth $W_n(j)>0$ of another uniformly chosen agent $j$ \label{r3}
	\end{enumerate}
	
	We can think of \ref{r1} as the agent losing a random proportion of wealth, \ref{r2} as no change in the agent's wealth and \ref{r3} as the agent being removed from the system and being replaced uniformly with another agent with positive wealth. We note that \ref{r3} is a simple approximation to resampling the agent's wealth from the current wealth distribution. 
	We focus here on simulations with the more realistic compromise mechanism \ref{r1}. In Appendix \ref{ap_gen_init} we will present simulation results for the more extreme replacement mechanisms \ref{r2} and \ref{r3} which lead to similar results, confirming that our model is not very sensitive on the choice of the replacement mechanism.

\begin{figure}[H]
		\centering
		\captionsetup{justification=centering}
		\includegraphics[width=\textwidth]{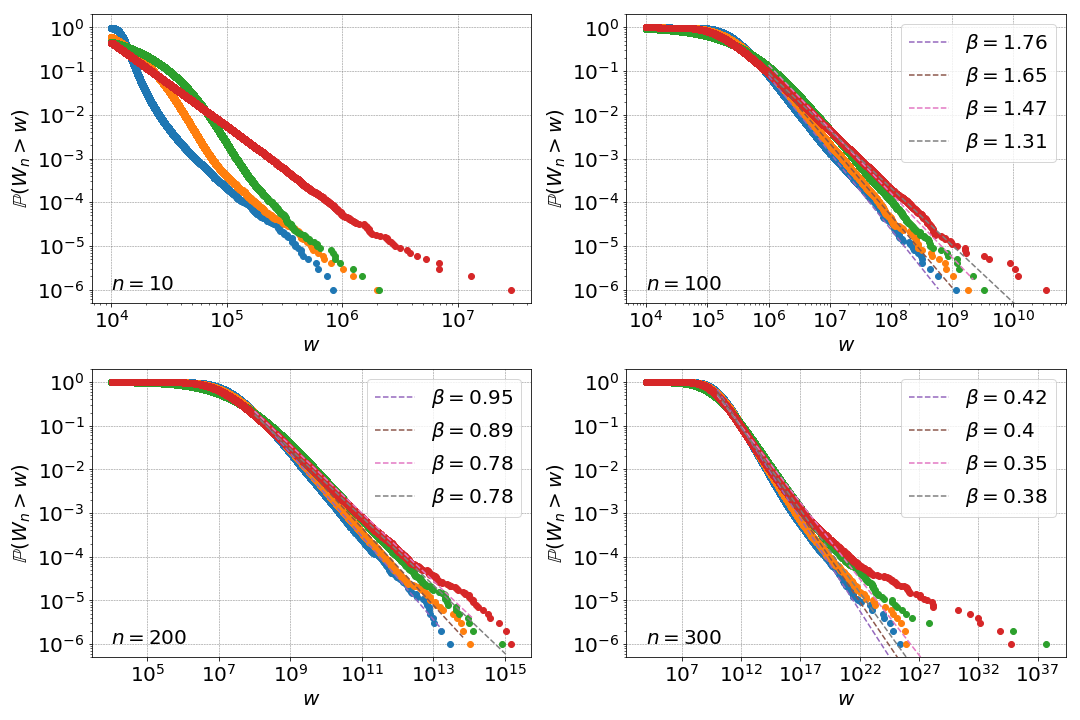}
		\caption{Empirical tails for simulation (\ref{nlkp}) with $N=10^6$ agents, zero savings $S_n=0$, $\alpha_n \sim \text{nct}(k,c,l,s)$ with $\gamma=1.075$, fitted parameters in \eqref{params}, the four initial conditions \ref{init1}-\ref{init4} with colour code, and replacement mechanism \ref{r1}. Power law fits show heavier tails with exponents $\beta$ decreasing with increasing times $n=10,100,200$ and $300$. 
		}
		\label{nct_nl_kesten_bank3_g1}
	\end{figure}

For each initial distribution we run the simulations iteratively using (\ref{nlkp}) for $N=10^6$ independent agents and \textbf{zero savings} $S_n=0$ with parameters in \eqref{params} and replacement mechanism \ref{r1}. We choose zero savings for convenience in this section, to isolate the effect of the multiplicative dynamics which is 
dominant in generating the wealth distribution in this model, see Appendix \ref{ap_nlkp}.
Results for empirical tail distributions at times $n=10,100,200$ and $300$ are presented in Figure \ref{nct_nl_kesten_bank3_g1}, using the colour code indicated in  \ref{init1}-\ref{init4}.
We also show standard inequality measures (see Appendix \ref{sec_ineq} for the definitions), the Gini coefficient $g$ and the top one percent income share $s_{0.01}$ for $\gamma=1.075$ up to time $n=300$ in top left and right of Figure \ref{nct_nl_kesten_bank3_g1_3_gini_s01}.
We see that all initial conditions eventually lead to monopoly, and for intermediate times power-law tails emerge in the wealth distribution. Due to the crossover \eqref{crossover} to super-exponetial growth, a two-tailed structure emerges for large times and wealth values.

\begin{figure}[H]
		\centering
		\captionsetup{justification=centering}
		\includegraphics[width=\textwidth]{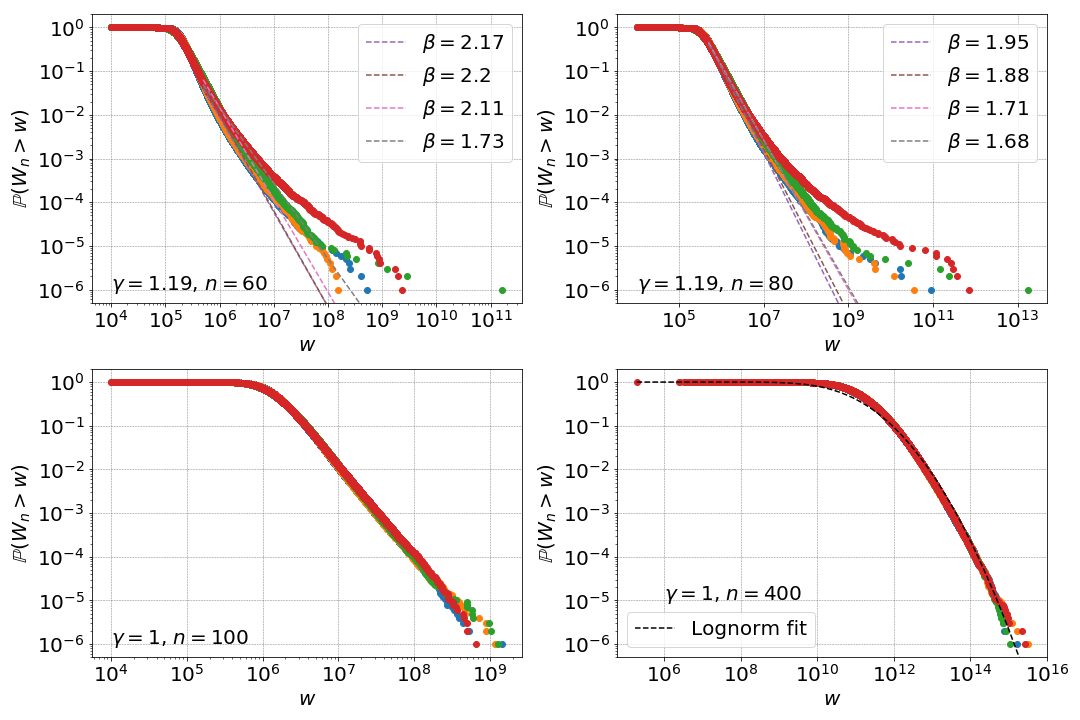}
		\caption{Top left and right: empirical tails for simulation (\ref{nlkp}) with $N=10^6$ agents, zero savings $S_n=0$, $\alpha_n \sim 0.23 \cdot \text{nct}(k,c,l,s)$ with fitted parameters \eqref{params} but with $\gamma=1.19$ for the four initial conditions with respective colour coding \ref{init1}-\ref{init4}, replacement mechanism \ref{r1} and power law fits with exponent $\beta$. 
		Bottom left and right: empirical tails for simulations as in top row, but with $\gamma=1$,  $S_n=0$, $\alpha_n \sim 2.5 \cdot \text{nct}(k,c,l,s)$, with lognormal fit at $n=400$. 
		}
		\label{nct_nl_kesten_bank3_g2_3}
	\end{figure}

\begin{figure}[H]
		\centering
		\captionsetup{justification=centering}
		\includegraphics[width=\textwidth]{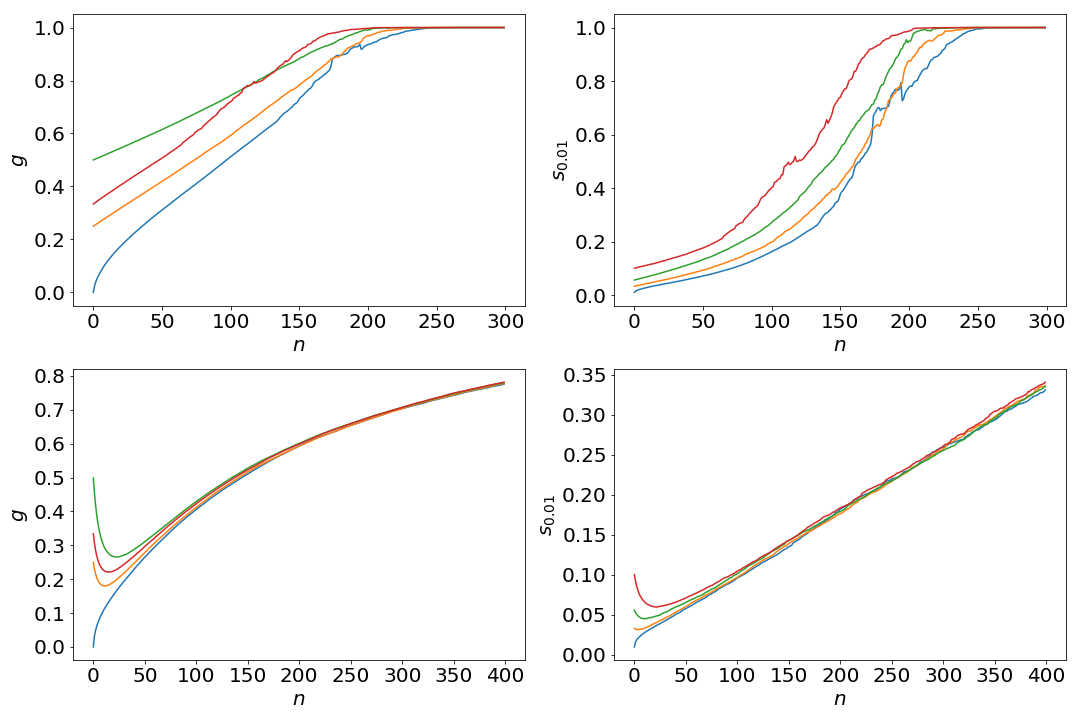}
		\caption{Gini, $g$, and top 1\% wealth shares, $s_{0.01}$, for simulation (\ref{nlkp}) with $N=10^6$ agents, zero savings $S_n=0$, $\alpha_n \sim \text{nct}(k,c,l,s)$ with fitted parameters in \eqref{params}, $\gamma=1.075$ for top left and right and $\alpha_n \sim 2.5 \cdot \text{nct}(k,c,l,s)$,  $\gamma=1$ for bottom left and right. The four initial conditions with respective colour coding \ref{init1}-\ref{init4} are used with replacement mechanism \ref{r1}. 
		}
		\label{nct_nl_kesten_bank3_g1_3_gini_s01}
	\end{figure}

\begin{figure}[H]
		\centering
		\captionsetup{justification=centering}
		\includegraphics[width=\textwidth]{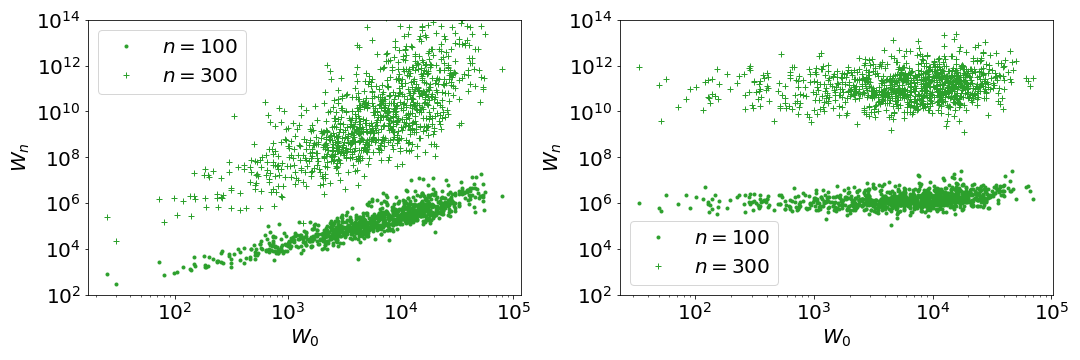}
		\caption{$W_0$ versus $W_n$ for $1000$ randomly chosen agents for simulation (\ref{nlkp}) with $N=10^6$ agents, zero savings $S_n=0$ with fitted parameters in \eqref{params}, left $\alpha_n \sim \text{nct}(k,c,l,s)$, $\gamma=1.075$ and right $\alpha_n \sim 2.5 \cdot \text{nct}(k,c,l,s)$ and $\gamma=1$. We use initial conditions with colour coding \ref{init3} and replacement mechanism \ref{r1}. We see a clear dependence on initial conditions for $\gamma >1$, and essentially no dependence for $\gamma =1$.
		}
		\label{nct_nl_kesten_bank3_g1_3_w0_wn}
	\end{figure}

	In Figure \ref{nct_nl_kesten_bank3_g2_3} we show for comparison empirical tails for $\gamma=1.19$ with
	 $\alpha_n \sim 0.23 \cdot \text{nct}(k,c,l,s)$, 
	 and for $\gamma=1$ with $\alpha_n \sim 2.5 \cdot \text{nct}(k,c,l,s)$, so that average ROR values are well approximated as shown in Figure  \ref{returns_all2}. For $\gamma=1$ we also compute the two inequality measures $g$ and $s_{0.01}$ up to $n=400$, see bottom left and right of Figure \ref{nct_nl_kesten_bank3_g1_3_gini_s01} which shows the independence of initial conditions and slower progression towards monopoly.  For the higher value of $\gamma$ we see that the crossover sets in earlier at more realistic wealth values around $10^7$ with a two-tailed structure with quite realistic power-law tails (cf.\ Figure \ref{survey_times_forbes2}). For the linear model with $\gamma =1$ we see no crossover and can fit the distribution for large times well by a log-normal distribution in accordance with \eqref{linres}. In this case there is also no noticeable difference between distributions originating from different initial conditions as we have seen in Figure \ref{nct_nl_kesten_bank3_g1_3_gini_s01}. This is also illustrated in Figure \ref{nct_nl_kesten_bank3_g1_3_w0_wn}, where we also see a clear dependence of final wealth values on initial conditions in the non-linear case with $\gamma >1$.

	\subsection{Realistic Initial Conditions}
	\label{sec_real_init_condns}
\begin{figure}[H]
		\centering
		\captionsetup{justification=centering}
		\includegraphics[scale=0.35]{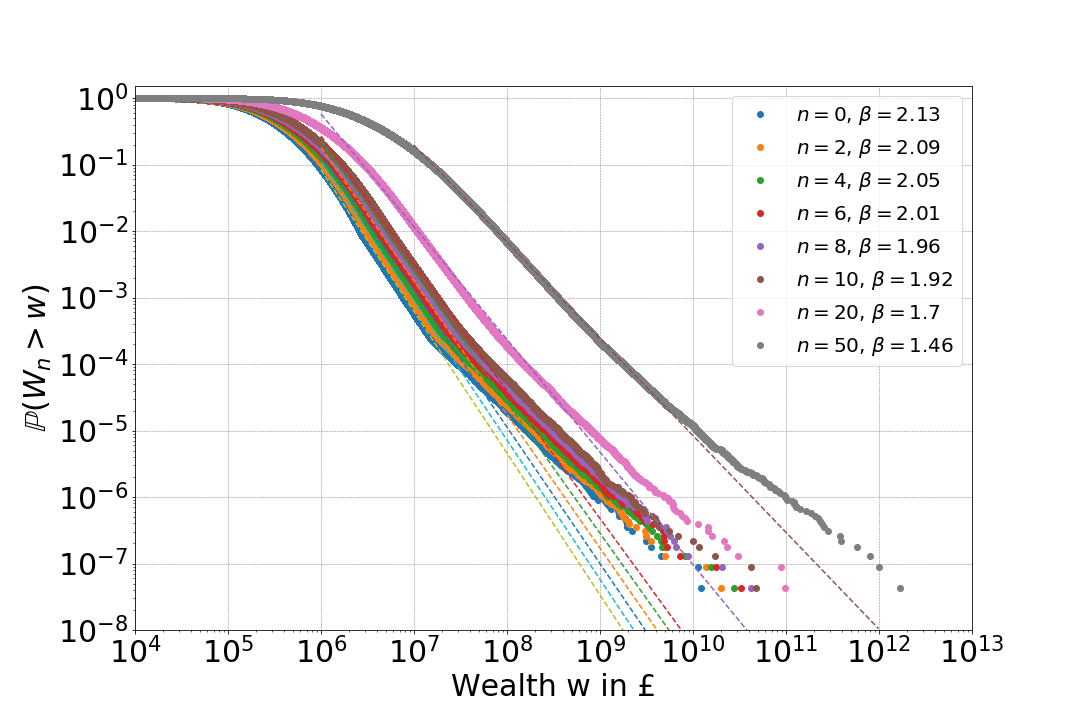}
		\caption{Empirical tails for simulation (\ref{nlkp}) with $N \approx 23 \cdot 10^6$ agents, replacement mechanism \ref{r1}, $\gamma=1.075$, fixed savings $S_n=S(W_0)$ \eqref{snfit}, $\alpha_n \sim \text{nct}(k,c,l,s)$ with fitted parameters in \eqref{params} for 2008 initial conditions. Fit values for a power-law tail exponent $\beta$ decrease from the initial value $2.13$.
		}
		\label{nct_nl_kesten_pop3}
	\end{figure}

	In this section we simulate a realistic scenario for the UK, with $N = 23 \cdot 10^6$ households, initial conditions $W_0$ extracted from the UK wealth distribution in 2008, and with fixed savings $S_n=S(W_0 )$ as given in \eqref{snfit} of Section \ref{savings}. Figure \ref{nct_nl_kesten_pop3} shows the empirical tail of the resulting wealth distribution
	at times $n=0,2,4,6,8,10,20$ and $50$, after simulating (\ref{nlkp}) with $S_n=S(W_0 )$, $\gamma=1.075$, $\alpha_n \sim \text{nct}(k,c,l,s)$ with fitted parameters in \eqref{params} and replacement mechanism \ref{r1}. Figure \ref{nct_nl_kesten_pop4}, in Appendix \ref{ap_real_init}, shows empirical tails for the other two replacement mechanisms \ref{r2}, \ref{r3} which lead to very similar results. The number of agents ($N$) is a rough estimate for the number of households in the UK with positive wealth in 2016. Time $n$ corresponds to the number of years after 2008, so for example $n=8$ corresponds to 2016. Again we can see increasing inequality, see Figure \ref{nct_nl_kesten_g_ws} in Appendix \ref{ap_real_init}, with the decreasing power-law exponent $\beta$.
	
	
	\begin{figure}[H]
		\centering
		\captionsetup{justification=centering}
		\includegraphics[scale=0.35]{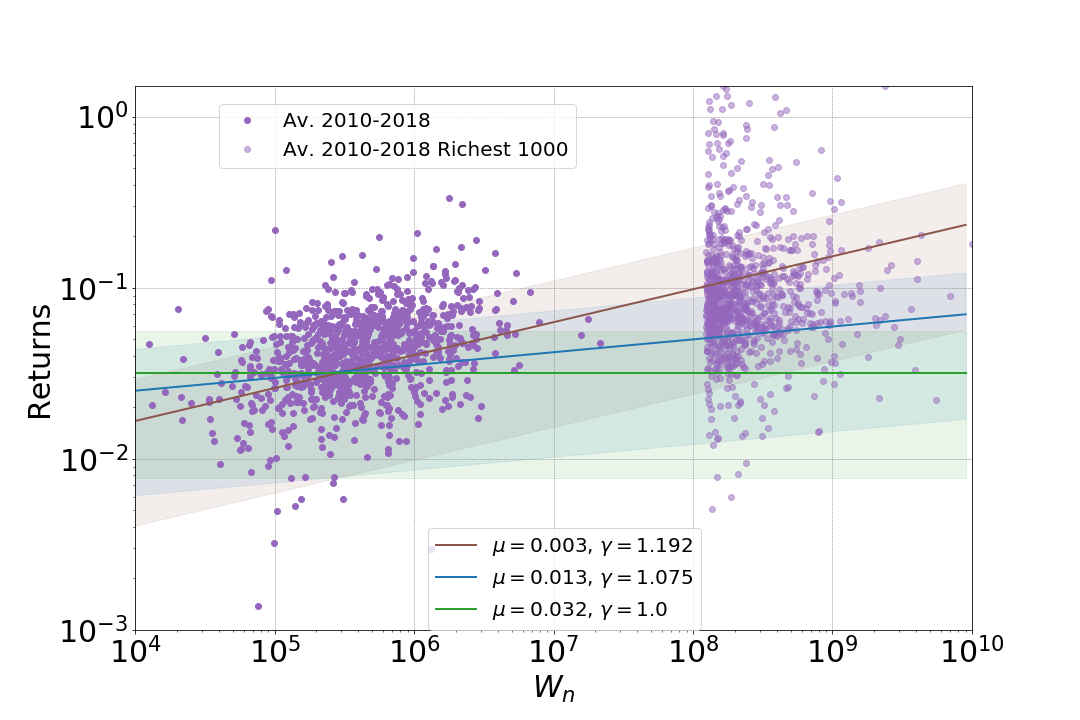}
		\caption{Average return over 2010-2018 for randomly chosen agents against agents wealth in 2018. 
			The  power fits (straight lines) for $\mathbb{E}[R_{n+1}|W_n]=\mu W_n^{\gamma-1}$, are the three	
			fits to the real world data from Figure \ref{returns_all2}, along with one standard deviation error region.}
		\label{nct_nl_realin_av_returns}
	\end{figure}
	
	Comparing Figures \ref{nct_nl_kesten_pop3} to Figure \ref{survey_times_forbes2} we see that the two-tailed structures differ slightly: While the heavier tail for billionaires with a power-law exponent of about $\beta=1$ is shifting but well preserved, the stability of the lighter power-law tail for millionaires is not well represented in our simulation. This is because we deliberately chose a simple model assuming that average ROR follows a monotone power law with wealth. While this is largely consistent with data, the survey data for RORs show some plateau  behaviour for millionaires clearly visible in Figure \ref{returns_all2}, which has also been suggested for other countries, see Figure 2 of \cite{ederer2020rich}. This may be related to the changing wealth composition of the very rich \cite{Visual_capitalist}.

	\newpage
	
	\section{Discussion}\label{sec_discussion}

	The model defined by the iterative equation (\ref{initw_eq}) represents a generic evolution of household wealth, based on the well motivated assumption that wealth exchange between households does not play an important role. The particular form \eqref{nlkp} of a non-linear Kesten process has been motivated by inferring empirically that RORs increase with household wealth, and that this relationship is consistent with a simple power law with exponent $\gamma$ as in (\ref{R_n}), see also Figure \ref{returns_all2}. We want to stress that the qualitative results and main features of our model do not depend on this particular choice, which we have taken for simplicity and in order to study the effect of the non-linearity with a single parameter. 
	We have seen from theory and simulations that the asymptotic dynamics of the model (\ref{nlkp}) and the resulting tail of the wealth distribution is dominated by the exponent $\gamma$. 
	For the linear case with $\gamma=1$ the RORs do not depend on wealth, and it is known that wealth grows asymptotically with a lognormal distribution (see Appendix \ref{ap_lkp}), which does not correspond to power-law tails seen in real data as in Figure \ref{survey_times_forbes2}. As demonstrated by our main results, the non-linear model with $\gamma >1$ exhibits power-law tails from generic initial conditions, including even perfect equality or light tailed exponential distributions, see Section \ref{sec_gen_sim}. It also leads to a two-tailed structure resulting from a crossover (\ref{crossover}) to super-exponential growth for the richest households.
	
We now summarise the most important theoretical features and differences of the linear ($\gamma =1$) and the non-linear ($\gamma >1$) non-stationary Kesten process \eqref{nlkp}:	
\begin{itemize}
    \item for all $\gamma\geq 1$, including the linear case, the model exhibits \textbf{monopoly}, i.e.\ for $N$ independent households the wealth fraction of the richest household increases with time and asymptotically approaches $1$\footnote{nevertheless, realistic levels of inequality can of course be achieved on intermediate timescales};
\item the linear model is \textbf{ergodic}, in the sense that the asymptotic exponential growth rate of household wealth does not depend on the initial condition $W_0$. The latter only enters as a multiplicative factor and the model is \textbf{scale invariant}, i.e.\ wealth can be measured in units of $W_0$ in a dimensionless way;
    \item the non-linear model is \textbf{not ergodic}, i.e.\ the asymptotic exponential growth rate depends on $W_0$ and the early dynamics. It is also \textbf{not scale invariant}, and the non-linearity on the right hand side leads to a \textbf{critical scale} \eqref{crossover} where wealth gain per year can exceed current wealth, which is observed in data for the richest households.

\end{itemize}
Moreover, we would like to stress that our model is phenomenological and not built from first principles, since we simply assume an empirically motivated non-linear relationship between ROR and current wealth. Therefore the model lacks a natural scale invariance and the parameter $\alpha_n$ is not universal, but depends on the units of measurement (the currency) and will vary between different countries/economic areas. On the other hand, the non-linearity induces a crossover scale that can be a possible explanation for an apparent two-tailed structure in the data. This is an important aspect of our model which should be investigated further. While not present in data from the USA, the two-tailed structure has been observed \cite{vermeulen2018fat} for several countries which have a less liberal economic system and put more emphasis on social equality. Related political measures such as taxation then lead to a more even wealth distribution and a lighter power-law tail for rich households including millionaires, while the richest in society distribute their wealth globally and can escape such measures, leading to a heavier tail for billionaires.

Other interesting generalisations to make the model more realistic include dynamics for negative wealth, a realistic treatment of bankruptcy events and also household lifetime and fragmentation over longer time periods, or a household dependence of the parameter $\alpha_n$ reflecting variations in ``fitness" to generate returns from investment. Also, mechanisms of household interaction possibly via a general redistribution or taxation procedure could be included and could lead to interesting effects on the dynamics similar to recent work in \cite{berman2020wealth}. But the aim of this paper was to introduce a simple model, that can explain the main features of wealth distribution and dynamics, and how they can be explained by a non-linear wealth dependence of the rate of return.

	
	\addcontentsline{toc}{section}{References}
	
	{\raggedright
 	\bibliography{mybib.bib}
 	\bibliographystyle{chicago}
    }
	
	\newpage

	\begin{appendices}
		

		\section{Theory}
		
		\subsection{Mean and Variance of Returns}
		
		\label{ap_mean_var_ret}
		
		With $R_n$ defined as in (\ref{R_n}) and $\alpha_n$ i.i.d. from some distribution with $\mu=\mathbb{E}[\alpha_{n+1}]$ and $\sigma^2=\text{var}(\alpha_{n+1})$ as in Section \ref{sec_model} we have
		\begin{align}
		\mathbb{E}[R_{n+1}|W_n] &= \mathbb{E}[\alpha_{n+1}W_n^{\gamma-1}] =\mathbb{E}[\alpha_{n+1}]W_n^{\gamma-1} = \mu W_n^{\gamma-1}, \label{exp_r}\\
		\text{var}(R_{n+1}|W_n) &= \text{var}(\alpha_{n+1}W_n^{\gamma-1}) = \text{var}(\alpha_{n+1})W_n^{2 (\gamma-1)} =  \frac{\sigma^2}{\mu^2}\mathbb{E}[R_{n+1}|W_n]^2 \label{var_r}.
		\end{align}
		Thus the interval $I$ of one standard deviation around the mean of $R_{n+1}$ used in Figure \ref{returns_all2} is
		
		\begin{equation}
		I=\left(\left(1-\frac{\sigma}{\mu}\right)\mathbb{E}[R_{n+1}|W_n],
		\left(1+\frac{\sigma}{\mu}\right)\mathbb{E}[R_{n+1}|W_n]\right) =
		((\mu-\sigma)W_n^{\gamma-1},(\mu+\sigma)W_n^{\gamma-1}).
		\label{one_std_mean}
		\end{equation}
		
\subsection{Linear Kesten Process}
		\label{ap_lkp}
		
		Consider a general linear Kesten process
		\begin{equation}
		W_{n+1}=A_{n+1}W_n+B_{n+1}\ , \quad n =0,1,2,\ldots\ ,
		\label{lin_Kesten}
		\end{equation}
		so that we can write
		\begin{equation}
		W_n = \prod_{k=1}^{n}A_k\left(W_0+\sum_{k=1}^{n}B_k \prod_{i=1}^{k}A_i^{-1}\right) .
		\label{W_n}
		\end{equation}
		Here $(A_n )_n$ and $(B_n )_n$ are sequences of i.i.d. random variables with
		\begin{equation}
		\mu := \mathbb{E}\big[ \log |A_k|\big] \in\mathbb{R}\quad\mbox{and}\quad
		\nu^2 := \text{Var}\big[ |\log A_k|\big]\in (0,\infty ).
		\end{equation}

\textbf{$\mu<0$}. For the stationary case, Kesten proved the following result, exact details can be found in Theorem 5 in \cite{kesten1973random}: if $\mu<0$, there exists $\beta>0$ such that $\mathbb{E}[|A_n|^{\beta}]=1$ and provided several other mild regularity conditions on the distributions of $A_n$ and $B_n$ are satisfied,
		\begin{equation}
		W_n \to W_{\infty} :=\sum_{k=1}^{\infty}B_k \prod_{i=1}^{k-1}A_i\quad\mbox{in distribution as }n\to\infty,
		\label{stat_w_dist}
		\end{equation}
for all initial conditions $W_0$. The stationary distribution exhibits a power law in one or both tails with parameter $\beta$, i.e. the following limits
		\begin{equation}
		\lim_{w \rightarrow \infty } w^{\beta}\mathbb{P}(W_{\infty}>w) \quad\text{and}\quad
		\lim_{w \rightarrow \infty } w^{\beta}\mathbb{P}(W_{\infty}<-w), \label{kl_limits}
		\end{equation}
exist and are finite, with at least one of them strictly greater than zero.\\

\textbf{$\mu>0$}. Following recent results in \cite{hitczenko2011renorming}, this non-stationary case can be analysed as follows. Taking absolute values and logarithms in \eqref{W_n} we get
		\begin{equation}
		\frac{\log |W_n| - \mu n}{\sqrt{n}\nu} 
		= \frac{\sum\limits_{k=1}^{n}\log |A_k| -\mu n}{\sqrt{n}\nu} 
		+ \frac{\log\left(\left|W_0+\sum\limits_{k=1}^{n}B_k \prod\limits_{i=1}^{k}A_i^{-1}\right|\right)}{\sqrt{n}\nu}.
		\label{leading_order}
		\end{equation}
		By the CLT for i.i.d.\ random variables $A_k$ we have
		\begin{equation*}
		\frac{\sum\limits_{k=1}^{n}\log |A_k| -\mu n}{\sqrt{n}\nu} \to \mathcal{N}(0,1)\quad\mbox{in distribution as }n\to\infty.
		\end{equation*}
Since $\mathbb{E}[\log |A_i|^{-1}] =-\mu <0$, we have $\left|W_0+\sum\limits_{k=1}^{n}B_k \prod\limits_{i=1}^{k}A_i^{-1}\right| \xrightarrow[]{n \rightarrow \infty} \widetilde W_{\infty} < \infty$, corresponding to the limit in the stationary case. This implies $		\frac{1}{\sqrt{n}\nu}\log\left(\left|W_0+\sum\limits_{k=1}^{n}B_k \prod\limits_{i=1}^{k}A_i^{-1}\right|\right)\to 0$, and
		\begin{equation}
		\frac{\log |W_n| - \mu n}{\sqrt{n}\nu} \to Z\sim \mathcal{N}(0,1)\quad\mbox{in distribution as }n\to\infty.
		\label{lin_norm}
		\end{equation}
This implies that 
\begin{equation*}
\frac{|W_n|^{1/(\sqrt{n}\nu)}}{e^{\mu \sqrt{n}/\nu}} \xrightarrow[]{d} e^Z \sim \text{Lognorm} (0,1)\quad\mbox{in distribution as }n\to\infty.
		\end{equation*}
Using the natural scale invariance of \eqref{W_n} we get a linear dependence on the initial condition, and to leading exponential order \eqref{leading_order} as $n\to\infty$
		\begin{equation*}
		|W_n| \simeq |W_0|\exp(\mu n + \sqrt{n} \nu Z) .
		\end{equation*}
%
We note that \cite{hitczenko2011renorming} also includes the case with $\mu=0$ which we do not discuss here.

		\subsection{Non-Linear Kesten Process}
		\label{ap_nlkp}
		
We analyse the non-linear process with $\gamma >1$ given in \eqref{nlkp} as
$$
W_{n+1} = W_n +\alpha_{n+1} W_n^\gamma +S_{n+1}.
$$
For simplicity we assume $\alpha_n >0$, which implies that $W_n$ is increasing and strictly positive for all $n\geq 0$. Negative values of $\alpha_n$ will lead to bankruptcy events as $n\to\infty$, for which we apply replacement mechanisms \eqref{r1}-\eqref{r3} as explained in Section \ref{sec_sim}. Taking logarithms leads to
\begin{align}
		\log W_{n+1} &= \log \big( W_n+\alpha_{n+1}W_n^{\gamma}+S_{n+1}\big) \nonumber \\
		&= \gamma \log W_n + \log\big( \alpha_{n+1}+1/W_n^{\gamma-1}+S_{n+1}/W_n^{\gamma}\big) \nonumber\\
\text{so that}\quad X_{n+1} &= \gamma X_n + B_{n+1}, \label{X_n1}
		\end{align}
where $X_n := \log W_n$ and  $B_{n+1} :=\log(\alpha_{n+1}+1/W_n^{\gamma-1}+S_{n+1}/W_n^{\gamma})$. 
Now using (\ref{W_n}) we get
\begin{equation}
X_n = \gamma^n \Big( X_0+ \sum_{k=1}^{n}B_k \gamma^{-k} \Big)\quad\mbox{so that}\quad
\frac{X_n}{\gamma^n} \xrightarrow[]{d} X_0+ D \quad\text{as } n \rightarrow \infty,
		\label{X_n_gamma_converge}
\end{equation}
where $D:=\sum\limits_{k=1}^{\infty} B_k \gamma^{-k}$. Since $W_n >0$ is increasing with $n$ and $\alpha_n$ are i.i.d., $B_k$ are bounded random variables, so $D\in (0,\infty )$ is a well defined random variable since $\gamma >1$.  Thus, as $n\to\infty$, this implies to leading exponential order
		\begin{equation}
		\frac{X_n}{\gamma^n} \simeq X_0+ D \quad\mbox{so that}\quad W_n \simeq \left(W_0 e^D \right)^{\gamma^n}.
		\label{W_n_scale}
		\end{equation}

		\section{Empirical Analysis}
		
		\subsection{Data Sources}
		
		\label{sec_data}
		
		Here we list the data sources used in the paper:
		
		\begin{enumerate}
			\item Biannual wealth and asset survey data 2008-2016 from the Office for National Statistics (ONS) \cite{was}
			\item Forbes rich lists \cite{Forbes_rl}
			\item Times rich list data - extracted from Times online newspaper 2019, 2020. For 2020 see \cite{Times2020_rl}. 2019 is no longer easily available online, please contact author.
			\item ONS household income, salary and expenditure data
			\cite{ONS_income, ONS_expenditure}
		\end{enumerate}
		
		\subsection{Notation for Tails and Power Laws}
		
		\label{sec_tail_pl}
		
A random variable $X\geq 0$ exhibits a power-law tail with parameters $\alpha ,\beta >0$ if		
		\begin{equation*}
		\mathbb{P}(X>x) \simeq \frac{\alpha}{x^{\beta}} \quad\text{to leading order as } x\to\infty,
		\end{equation*}
where $\beta$ is called the power-law exponent. 
The Pareto distribution is a standard example for power-law tails, where
		\begin{equation*}
		X \sim \text{Pareto}(x_m,\beta)\quad\mbox{if}\quad \mathbb{P}(X>x)=\Big(\frac{x_m}{x}\Big)^\beta \quad\mbox{for }x\geq x_m.
		\end{equation*}
We can estimate the tail of a random variable $X$ by the empirical tail $\mathbb{P}_N(X>x)$ of a random sample $\{x_1,x_2, \dots, x_N\}$ of $X$ defined as
\begin{equation*}
		\mathbb{P}_N(X>x) := \frac{1}{N} \sum_{i=1}^{N}\mathbf{1}_{x_i>x}
		\end{equation*}
where $\mathbf{1}$ is the indicator function. It is a standard result that 
		$\mathbb{P}_N(X>x) \rightarrow \mathbb{P}(X>x)$ almost surely as $N \rightarrow \infty$.
		
To fit the power-law tail parameters $\alpha$ and $\beta$ we use linear regression in a window of interest on the double logarithmic scale $\log \mathbb{P}_N(X>x)$ vs. $\log x$. This simple technique is known to introduce a bias in the fit \cite{Newman}, but for our purposes in this paper we find it is a sufficient approximation.
		

\subsection{Tail of UK Wealth}
		
\label{sec_tail_UK_wealth}
		
		Here we outline how we extract the empirical tail from wealth survey and rich list data from \cite{was, Forbes_rl}. For extensive discussion on the wealth and asset survey see \cite{ONS_wealth_2008}. We have wealth survey data in the form $(\tilde{h}_i, \tilde{w}_i)$ for $i=1,2, \dots n$ where
		$\tilde{h}_i \in [0,1]$ is the cumulative proportion of households and $\tilde{w}_i \in [0,1]$ is their corresponding cumulative proportion of wealth. Let the data be ordered with increasing wealth per household, 
then the 
		the Gini coefficient can be calculated from the Lorenz curve \cite{was} defined by the points $(\tilde{h}_i, \tilde{w}_i)$. Let $H$ and $W$ be the total number of households and the total amount of wealth of all households respectively.
		
		Define $\hat{w}_i :=(\tilde{w}_{i+1}-\tilde{w}_{i})W$ and $\hat{h}_i :=(\tilde{h}_{i+1}-\tilde{h}_{i})H$ for $i=1,2, \dots, n-1$. Then $\hat{w}_i$ is the total amount of wealth owned by an increasingly rich number $\hat{h}_i$ of households, so that 
		$w_i :=\dfrac{\hat{w}_i}{\hat{h}_i}$ is the corresponding average amount. Since the original data was ordered we have also $w_i \leq w_{i+1}$ for all $i$. 
		Therefore the points $(w_i,\tilde{h}_{i+1})$ characterise an approximation to the empirical CDF and the points $(w_i,1-\tilde{h}_{i+1})$ give
		the corresponding approximation to the empirical tail of the wealth distribution. We plot the empirical tail of positive wealth in Figure \ref{survey_times_forbes2} which are points below \pounds$10^8$ for the years 2008, 2010, 2012, 2014 and 2016.
		
		We have separate data in the form of rich lists for individual household wealth $w_i$ for $i=1,2,\dots,R$, where $R$ are the number of households in the rich list. We assume the rich list contains the $R$ wealthiest households in the population, so for ordered $w_i$ the empirical tail of their wealth distribution is given by the points and $(w_i,(R-i)/H)$. The empirical tail of the rich lists are the points above \pounds$10^8$ in Figure \ref{survey_times_forbes2}, and are matched in colour for corresponding years to the survey data.
		

		\subsection{Estimating the ROR}
		
		\label{sec_returns}
		
To approximate returns of individual household wealth using the survey data \cite{was} we use the returns on percentile wealth. For each time period $n$ from the empirical tail of the survey we extract the percentile $w_{i,n}$ such that 
$\mathbb{P}_N (W_n >w_{i,n})=p_i$ where $p_i=1-i/100$ for $i \in \{1,2,\dots,100\}$. Note we only extract positive percentiles $i$ such that $w_{i,n}>0$, which excludes the poorest households. 
Then we substitute $W_n=w_{i,n}$ in (\ref{returns_eq}) to calculate RORs of percentiles over each of the five biannual time periods 2008-2016 of the data \cite{was}. For percentile $i$ we have the ROR as
\begin{equation}
    	r_{i,n+2} = \frac{w_{i,n+2}-w_{i,n}-2s_{i,n+2}}{2w_{i,n}},
 \label{wealth_growth_approx}
\end{equation}
where $s_{i,n}$ are the savings in percentile $i$ and time period $n$ (see Section \ref{savings} for details).
		
Note that these percentile RORs for the ONS survey data \cite{was} plotted in Figure \ref{returns_all2} only approximate RORs for individual households. Our procedure does not account for households changing percentiles over a time period, leading to reduced fluctuations of the resulting returns data. To compensate for this and also possible effects of the financial crisis from 2008 onwards, we combine all time periods in a single data set to infer system parameters.
For billionaires we have individual wealth data across time. We ignore savings to compute returns according to (\ref{returns_eq}), and plot these values for 2016 in Figure \ref{returns_all2}.
		
\begin{figure}[H]
	\centering
	\captionsetup{justification=centering}
	\includegraphics[width=\textwidth]{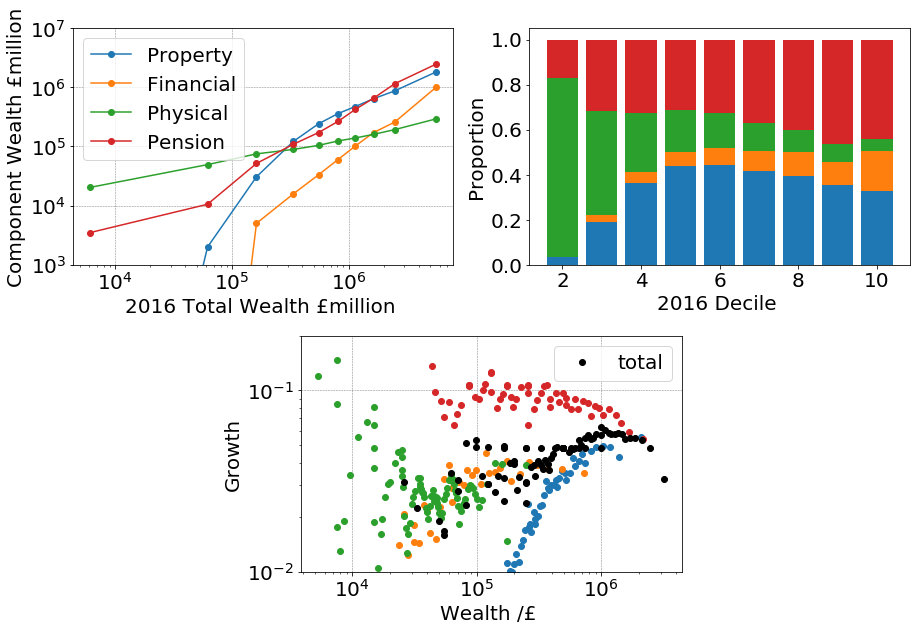}
			\caption{Absolute wealth by components as a function of total wealth (top left) and wealth proportions by component of positive wealth deciles (top right), both from ONS data \cite{was} from 2016. Bottom: ROR with zero savings (wealth growth) and same colour code averaged over time periods from 2008 to 2016, computed as described in \eqref{wealth_growth_approx} from percentile data \cite{was}.}
			\label{component_wealth_growth_prop1}
		\end{figure}

		
In order to understand the dependence of ROR on wealth in particular for the UK, it is instructive to consider the different composition of wealth for poorer and richer households. Survey data \cite{was} differentiate four components of wealth: property, physical, financial and pension, and their typical distribution is summarized in Figure \ref{component_wealth_growth_prop1}, exemplary for 2016 data. Financial and property wealth of the poorest decile have a negative sign (i.e.\ constitute debt), and the total average wealth in that decile is approximately $0$ and not shown in Figure \ref{component_wealth_growth_prop1}.


		

		The paper `The rate of return on everything, 1870--2015' \cite{jorda2019rate} provides a comprehensive analysis of average returns across four different types: bills, bonds, equity and housing over 1870-2015.  In particular for the period 1980-2015 the average real rate of returns on equity and housing for the UK are 9.11\% and 6.81\%, respectively (Table 7, p 37 \cite{jorda2019rate}). Therefore, the increasing proportion of property and financial wealth for wealthier households can account for RORs increasing with wealth. This is also confirmed in Figure \ref{component_wealth_growth_prop1} (bottom), where we see that ROR (technically ROR with zero savings as it is unclear how to divide savings across components) for physical and pension wealth are largely independent of wealth, while property and financial ROR increase with wealth.

		\subsection{Inequality Measures}
		\label{sec_ineq}
		There are several measures of inequality each with their various merits \cite{Champernowne_ineq,piketty2013capital}. In this paper we use the standard Gini coefficient $g \in [0,1]$  and top 1\% wealth share $s_{0.01} \in (0,1]$. The Gini coefficient can be thought of as a measure of the difference between any two randomly selected agents wealth. The top 1\% wealth share is defined as the proportion of wealth held by the richest 1\% of the population.

\begin{figure}[H]
			\centering
			\captionsetup{justification=centering}
			\includegraphics[scale=0.35]{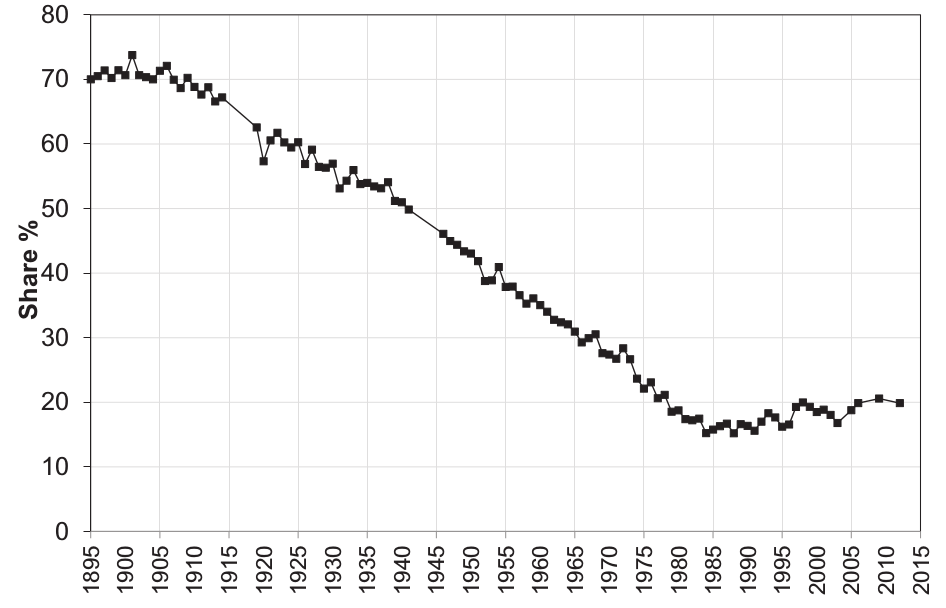}
			\caption{Top 1\% UK wealth share from 1895-2013 taken from \cite{alvaredo2018top}}
			\label{top1pc_ws_uk}
		\end{figure}

		For a non-decreasing ordered sample of $N$ agents' wealth 
		$w_1 \leq w_2 \leq \dots \leq w_N$ with total wealth $W=\sum\limits_{i=1}^N w_i$ we define
		
		\begin{equation}
		s_{0.01} := \sum\limits_{ i>0.99N}^{N}w_i \Big/ W \quad\mbox{and}\quad
		g := \frac{2}{N} \sum_{i=1}^{N}iw_i \Big/ W- \frac{N+1}{N} .		\end{equation}
We note the two extreme cases:
		\begin{enumerate}
			\item Perfect equality: $w_1=w_2=\dots=w_N \Rightarrow$ $g=0$ and $s_{0.01}=0.01$;
			\item Perfect inequality: $w_i =0$ for
			$i=1,2,\dots,N-1$ and $w_N >0$ $\Rightarrow$ $g=1$ and $s_{0.01} =1$.
		\end{enumerate}
The UK top one percent wealth share (see Figure \ref{top1pc_ws_uk}) has decreased significantly from 1895 until around 1985, and is increasing slightly since then.
		
		\subsection{Non-central $t$ Distribution}
		
		\label{sec_nct}
		
		We fit the coefficients $\alpha_n$ \eqref{alpha_n} with a shifted and scaled non-central $t$-distribution (nct) as seen Figure \ref{alpha_bil_was_kd}, which has been used for fitting stock returns that are both skewed and heavy tailed \cite{tsionas2002bayesian}. The standard non-central $t$ distribution  is defined by a random variable
		\begin{equation*}
		U=\frac{Z+c}{\sqrt{V/k}}\ ,\quad\text{where }Z \sim \mathcal{N}(0,1)\text{ is a standard Gaussian},
		\end{equation*}
		$c \in \mathbb{R}$ is the centrality parameter, and $V \sim \chi^2(k)$ with $k>0$ the degrees of freedom. The shifted and scaled non-central $t$-distribution we use is then given by the random variable
		\begin{equation*}
		sU+l \sim \text{nct}(k,c,l,s),
		\end{equation*}
		with the shift parameter $l\in \mathbb{R}$ and scale parameter $s >0$.

		\section{Supplementary Simulation Results}
		
		\subsection{Generic Initial Conditions}
		
		\label{ap_gen_init}
Recall the replacement mechanisms \eqref{r1}-\eqref{r3} in case of bankruptcy events:
	\begin{enumerate}[leftmargin=2cm, label=\textbf{R.\arabic*}]
		\item replace with a proportion of the agent's previous positive wealth value \newline $p W_{n-1}(i)>0$ such that $p$ is uniformly chosen from $(0,1]$ 
		\item replace with the agent's previous positive wealth value $W_{n-1}(i)>0$
		\item replace with wealth $W_n(j)>0$ of another uniformly chosen agent $j$ 
	\end{enumerate}
	
\begin{figure}[H]
		\centering
		\captionsetup{justification=centering}
		\includegraphics[width=\textwidth]{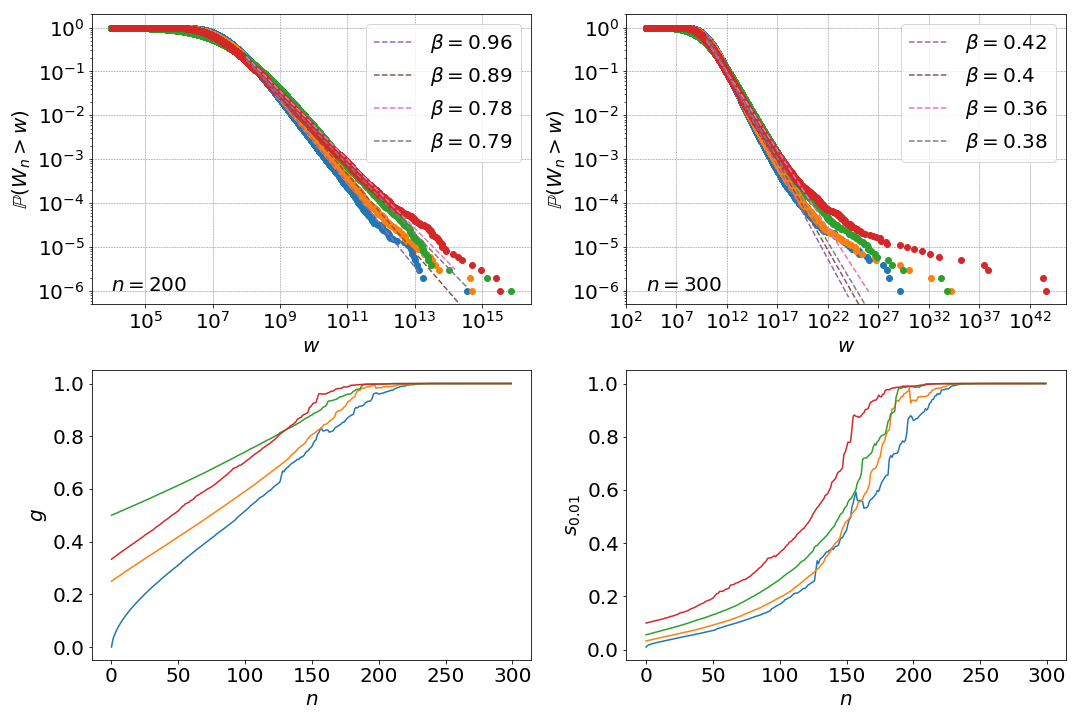}
		\caption{Simulation (\ref{nlkp}) with $N=10^6$ agents, zero savings $S_n=0$, $\alpha_n \sim \text{nct}(k,c,l,s)$ with fitted parameters in \eqref{params}  and $\gamma=1.075$ for the four initial conditions with respective colour coding \ref{init1}-\ref{init4} and replacement mechanism \ref{r2}. Top left and right show empirical tails at times $n=200,300$ and power-law tail fits with exponents $\beta$. 
		Bottom left and right show respective Gini, $g$, and top 1\% wealth shares, $s_{0.01}$ up to $n=300$.}
		\label{nct_nl_kesten_bank1_g1_gini_top1}
	\end{figure}

\begin{figure}[H]
		\centering
		\captionsetup{justification=centering}
		\includegraphics[width=\textwidth]{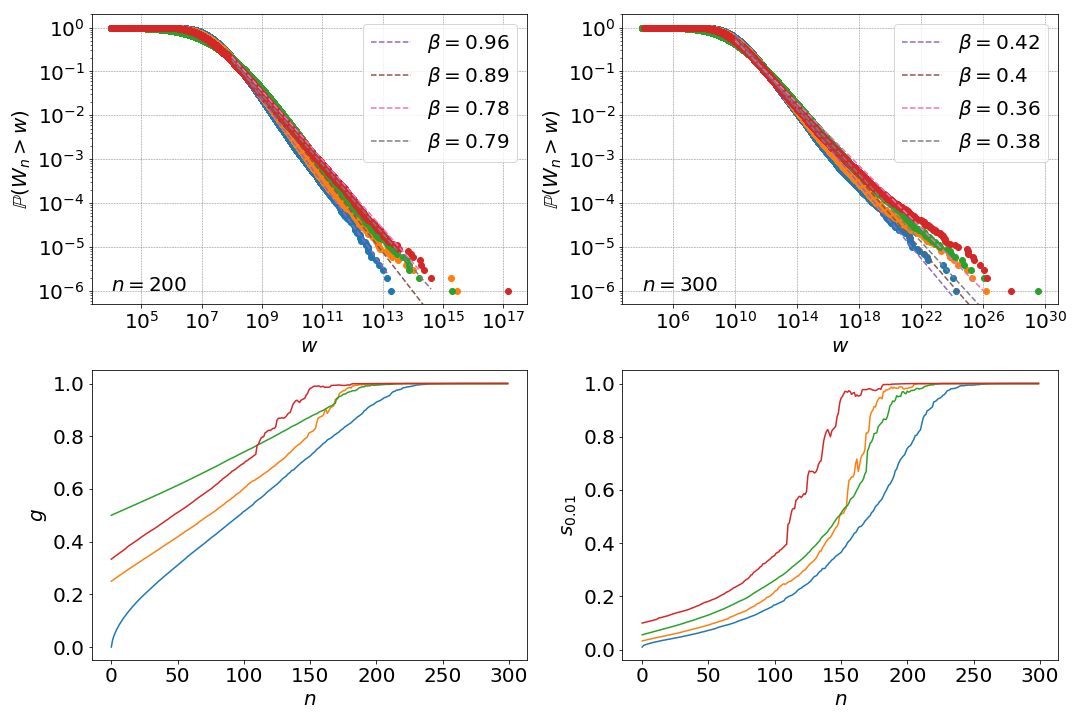}
		\caption{Simulation (\ref{nlkp}) with $N=10^6$ agents, zero savings $S_n=0$, $\alpha_n \sim \text{nct}(k,c,l,s)$ with fitted parameters in  \eqref{params}  and $\gamma=1.075$ for the four initial conditions with respective colour coding \ref{init1}-\ref{init4} and replacement mechanism \ref{r3}. Top left and right show empirical tails at times $n=200,300$ and power-law tail fits with exponents $\beta$. 
		Bottom left and right show respective Gini, $g$, and top 1\% wealth shares, $s_{0.01}$ up to $n=300$. }
		\label{nct_nl_kesten_bank2_g1_gini_top1}
	\end{figure}

We can see from Figures \ref{nct_nl_kesten_bank3_g1}, \ref{nct_nl_kesten_bank3_g1_3_w0_wn} (top left and right), \ref{nct_nl_kesten_bank1_g1_gini_top1} and \ref{nct_nl_kesten_bank2_g1_gini_top1} that the empirical tails and inequality measures of the simulations (\ref{nlkp}) evolve similarly in time for the three replacement mechanisms  \eqref{r1}-\eqref{r3} until the system enters the crossover region for wealth values around $10^{14}$ and time about $n=200$. Then bankruptcy events become more frequent and relevant for the richest households, leading to significant differences with mechanism \ref{r3} naturally leading to slowest growth.


		\subsection{Realistic Initial Conditions}
			\label{ap_real_init}
\begin{figure}[H]
			\centering
			\captionsetup{justification=centering}
\includegraphics[width=\textwidth]{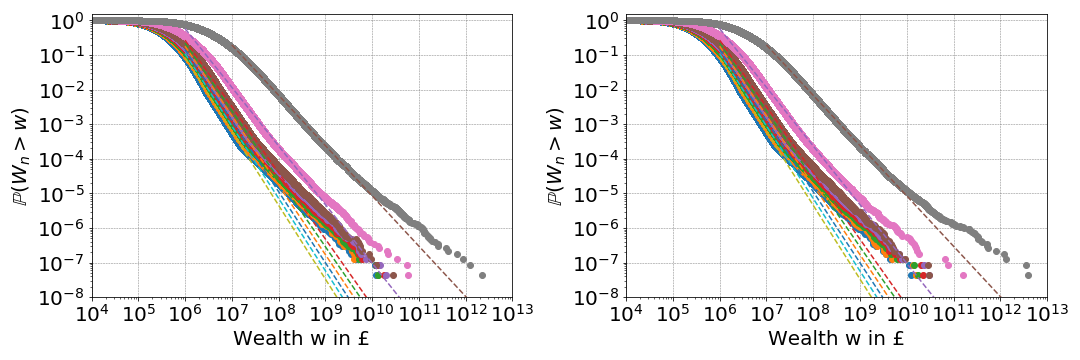}
			\caption{Empirical tails for simulation (\ref{nlkp}) with $N \approx 23 \cdot 10^6$ agents, replacement mechanisms \ref{r2} (left) and \ref{r3} (right), fixed savings $S_n=S(W_0)$ \eqref{snfit}, $\alpha_n \sim \text{nct}(k,c,l,s)$ with fitted parameters in \eqref{params} and $\gamma=1.075$ for 2008 initial conditions at times $n=0,2,4,6,8,10,20$ and $50$. Power law fits with exponents $\beta$ decreasing from $\beta=2.13$ at $n=0$ to $\beta=1.45$ at $n=50$.}
			\label{nct_nl_kesten_pop4}
		\end{figure}

\begin{figure}[H]
			\centering
			\captionsetup{justification=centering}
			\includegraphics[width=\textwidth]{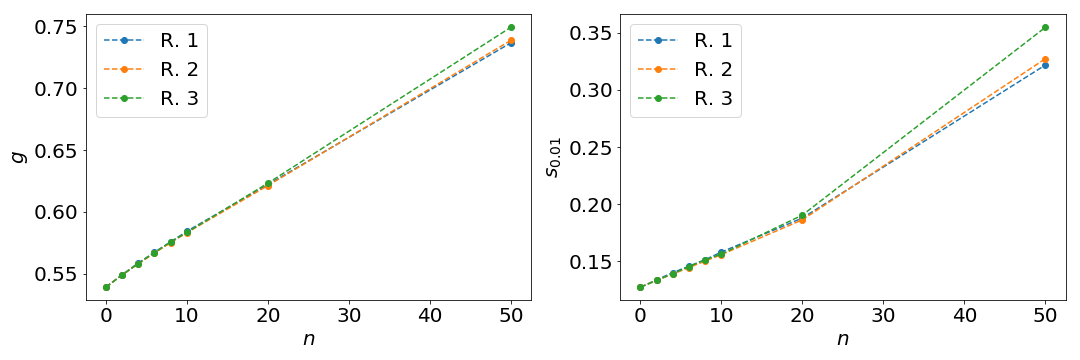}
			\caption{Gini, $g$, (left), top 1\% wealth shares, $s_{0.01}$, (right) for simulation (\ref{nlkp}) with $N \approx 23 \cdot 10^6$ agents, fixed savings $S_n=f(W_0)$, $\alpha_n \sim \text{nct}(k,c,l,s)$ with fitted parameters \eqref{params} and $\gamma=1.075$  with rough 2008 initial conditions and replacement mechanisms \ref{r1}-\ref{r3}. }
			\label{nct_nl_kesten_g_ws}
		\end{figure}

		We can see from Figures \ref{nct_nl_kesten_pop3},
		\ref{nct_nl_kesten_pop4} and \ref{nct_nl_kesten_g_ws} that the three replacement mechanisms \eqref{r1}-\eqref{r3} give very similar results on wealth distribution and inequality over time $n$, for the simulations described in the caption of Figure \ref{nct_nl_kesten_pop4}. This is due to the much shorter time horizon compared to numerical studies of generic initial conditions, and confirms that the choice of replacement mechanism is not crucial over limited time periods.
		

	\end{appendices}
	
\end{document}